\documentclass{egpubl}
\usepackage{eurovis2020-modified}

\SpecialIssuePaper         
\usepackage[T1]{fontenc}
\usepackage{dfadobe}  

\usepackage{cite}  
\BibtexOrBiblatex
\electronicVersion
\PrintedOrElectronic
\ifpdf \usepackage[pdftex]{graphicx} \pdfcompresslevel=9
\else \usepackage[dvips]{graphicx} \fi

\usepackage{egweblnk}
\usepackage{colonequals}
\usepackage{graphicx}
\graphicspath{{./figs/}}

\usepackage[dvipsnames]{xcolor}
\usepackage{color}

\usepackage{amssymb}
\usepackage{amsmath}
\usepackage{amsfonts}

\newcommand {\mm}[1] {\ifmmode{#1}\else{\mbox{\(#1\)}}\fi}

\newcommand{\Rspace}        {\mm{\mathbb{R}}}
\newcommand{\Xspace}        {\mm{\mathbb{X}}}
\newcommand{\Sspace}        {\mm{\mathbb{S}}}
\newcommand{\Mspace}        {\mm{\mathbb{M}}}

\newcommand{\TM}            {\mbox{T\Mspace}}

\newcommand{\tool}            {\mbox{{\sffamily MVF Designer}}}

\newtheorem{theorem}{Theorem}[section]
\newtheorem{corollary}{Corollary}[theorem]

\newcommand{\para}[1]        {\vspace{0mm}\noindent{\textbf{#1}}}

\usepackage{tikz}
\usetikzlibrary{calc}
\colorlet{dgren}{green!50!black}
\newcommand\minn[1]{
\draw[draw=blue,fill=white,thick] (#1) circle (.1);
\draw[draw=blue,fill=white,thick] (#1) circle (.05);
}
\newcommand\sadd[1]{
\draw[dgren,fill=white,thick] (#1) circle (.1);
\foreach \r in {0,90,180,270}{
  \draw[dgren,thick] (#1)--++(\r:.1);
}
}
\newcommand\maxx[1]{
\draw[draw=red,fill=white,thick] (#1) circle (.1);
\fill[fill=red] (#1) circle (.05);
}

\usepackage{mathtools}
\newcommand{\grad}[1]       {{\nabla {#1}}}

\title[MVF Designer]
     {MVF Designer: Design and Visualization of Morse Vector Fields}

\author[Y. Zhou et al.]
{\parbox{\textwidth}{\centering 
Youjia Zhou$^{1}$,
 Janis Lazovskis$^{2}$,
 Michael J. Catanzaro$^{3}$, 
 Matthew Zabka$^{4}$, 
 Bei Wang$^{1}$
}\\
{\parbox{\textwidth}{\centering 
$^1$ University of Utah, USA; e-mails: \{zhou325, beiwang\}@sci.utah.edu\\
$^2$ University of Aberdeen, UK; e-mail: janis.lazovskis@abdn.ac.uk\\
$^3$ Iowa State University, USA; e-mail: mjcatanz@iastate.edu\\
$^4$ Southwest Minnesota State University, USA; e-mail:mjzabka@ncsu.edu
}}}

\begin{document}

 \teaser{
 \vspace{-4mm}
  \includegraphics[width=0.95\linewidth]{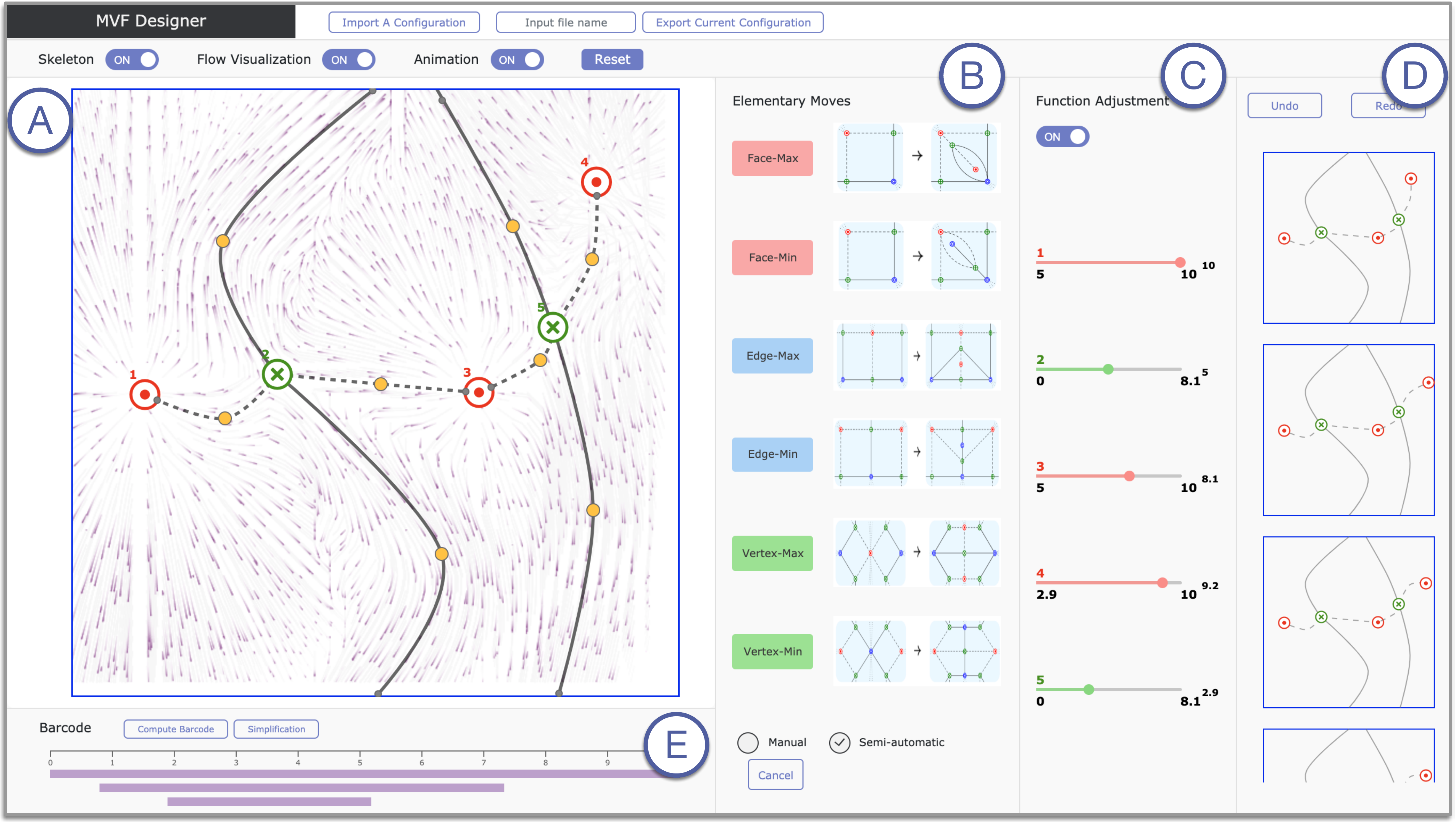}
  \centering
   \caption{With {\tool}, users can interact with and manipulate a Morse vector field at the structural and visual levels. The interface consists of several panels.
   (A) \textbf{Flow visualization} panel: Supports modifying the topology and geometry of the topological skeleton; visualizes the dynamics of the underlying vector field via animation.  
   (B) \textbf{Elementary moves} panel: Provides a set of elementary moves as fundamental building blocks of a vector field. 
   (C) \textbf{Function adjustment} panel: Allows modifying the function values at singularities.
    (D) \textbf{History} panel: Provides undo and redo features to remove or repeat single or multiple operations. 
    (E) \textbf{Barcode} panel: Computes and displays persistence barcodes to guide vector field simplification.}
 \label{fig:interface}
}

\maketitle

\begin{abstract}
Vector field design on surfaces was originally motivated by applications in graphics such as texture synthesis and rendering. 
In this paper, we consider the idea of vector field design with a new motivation from computational topology. 
We are interested in designing and visualizing vector fields to aid the study of Morse functions, Morse vector fields, and Morse--Smale complexes.  
To achieve such a goal, we present {\tool}, a new interactive design system that provides fine-grained control over vector field geometry, enables the editing of vector field topology, and supports a design process in a simple and efficient way using elementary moves, which are actions that initiate or advance our design process. 
Our system allows mathematicians to explore the complex configuration spaces of Morse functions, their gradients, and their associated Morse--Smale complexes. 
Understanding these spaces will help us expand further their applicability in topological data analysis and visualization.

\end{abstract} 


\section{Introduction}
\label{sec:introduction}

\para{Motivation from computational topology.}
The original motivation for vector field design on surfaces originates from diverse applications in graphics  (e.g.,~\cite{ZhangMischaikowTurk2006,ChenKwatraWei2012}). 
A \emph{designer vector field} can be used to define texture orientation and scale in example-based texture synthesis~\cite{PraunFinkelsteinHoppe2000, Turk2001, WeiLevoy2001}, to guide the orientation of brushes and hatches in non-photorealistic rendering~\cite{Hertzmann1998, HertzmannZorin2000}, and to simulate fluid flows on smooth surfaces of arbitrary topology~\cite{Stam2003}. 

Our work is newly motivated by advances in computational topology. 
In particular, our goal is to design and visualize vector fields to aid the characterization and classification of Morse functions, Morse vector fields, and Morse--Smale complexes.
All three concepts arise from Morse theory~\cite{Milnor1963}, which establishes the foundation for many techniques in topological data analysis and visualization.

Morse theory studies the relation between the shape of a space and functions on the space.  
It describes ``how the critical points of a function defined on a space affect the topological shape of the space, and conversely, how the shape of a space controls the distribution of the critical points of a function"~\cite{Matsumoto1997}. 
Formally,  given a Morse function $f\colon \Xspace \to \Rspace$ defined on a smooth manifold $\Xspace$, Morse theory associates the topological changes of the sub-level sets $\Xspace_a \colonequals f^{-1}(-\infty,a]$ with the critical points of $f$, as $a$ varies.  
The gradient of $f$ (with respect to some Riemannian metric on $\Xspace$) is a vector field consisting of vectors in the direction of the steepest ascent of $f$. 
Researchers in visualization have studied this vector field extensively~\cite{HeineLeitteHeike2016}. 
The Morse--Smale complex (MSC) captures the characteristics of this vector field by decomposing the manifold into cells of uniform flow~~\cite{BremerEdelsbrunnerHamann2004, EdelsbrunnerHarerZomorodian2003,
GyulassyNatarajanPascucci2005, GyulassyNatarajanPascucci2007}. The MSC is a type of topological summary that provides a compact and abstract representation of scalar functions, which has been shown to be effective for identifying, ordering, and selectively simplifying features of data across a wide range of applications (for example, in~\cite{EdelsbrunnerHarerNatarajan2003, EdelsbrunnerHarerZomorodian2003,Gyulassy2008,GyulassyBremerPascucci2008,GyulassyNatarajanPascucci2007,GyulassyKnollLau2015b,GyulassyKnollLau2015}).

Given the connections between Morse functions, their gradients, and Morse--Smale complexes, we envision a design tool with the ability to perform fine-grained analysis and exploration of these objects to provide broad applicability in the study of classic and discrete Morse theory and topological data analysis.

\para{Contributions.}
To this end, we present {\tool}, a powerful vector field design system that helps users characterize and classify the spaces of Morse functions, Morse vector fields, and Morse-Smale complexes.
In this paper, we discuss primarily the design of Morse vector fields; the design of Morse functions and MSCs comes (almost) for free. 
{\tool} supports not only visualization and graphics researchers, but also applied and computational topologists.
{\tool} will: 
\begin{enumerate}
\item Provide control over vector field topology, such as types and locations of singularities, and geometry of the separatrices;    
\item Generate a vector field with the exact topology as the user intended;
\item Enable topological algebra through editing of and computation with the vector field topology;
\item Support a design process in a simple and efficient way. 
\end{enumerate}
In addition, inspired by topological data analysis, {\tool} will encode and embrace the notion of persistence~\cite{EdelsbrunnerLetscherZomorodian2002}; it will:
\begin{enumerate}
\item  Compute and visualize the barcode~\cite{Ghrist2008} of a designer vector field to offer a global summary of its features; 
\item  Support the adjustment of function values at singularities to explore diverse Morse functions, their gradient vector field configurations and barcodes;
\item Create a one-to-one mapping between the topological features in the domain with bars in the barcode to guide interactive vector field simplification.  
\end{enumerate} 
With {\tool}, users can characterize and classify the spaces of Morse functions, Morse vector fields, and Morse--Smale complexes. In particular, they can:  
\begin{enumerate}
\item Construct and explore topological invariants in the classification of Morse--Smale flows on surfaces; 
\item Characterize various equivalence classes of Morse functions through persistence; 
\item Design and visualize Morse--Smale complexes; 
\item Study the inverse problem of generating Morse functions (and their gradients) from a given barcode; 
\item Explore combinatorics of vector fields.
\end{enumerate}
Finally, with {\tool}, users can create ensembles of Morse functions (Morse vector fields, or MSCs) for uncertainty quantification and visualization.

\section{Technical Background}
\label{sec:background}

We review the most relevant notions on vector fields, vector field topology, persistence,  Morse functions, and MSCs. See~\cite{PalisMelo1982} for an introduction to vector fields and~\cite{EdelsbrunnerHarer2008} for a survey on persistent homology. 
Throughout this paper, we restrict our attention to the design of vector fields on a sphere; although our framework are extendable to general orientable surfaces. 

\para{Vector field topology.} 
A 2D vector field (flow) $v$ is a smooth mapping $v \colon \Mspace \to \Rspace^2$ defined on a surface $\Mspace$ (i.e, a $2$-dimensional manifold). 
In this paper, we deal with smooth vector fields on a closed two-dimensional sphere $\Mspace \colonequals \Sspace^2$. 
Although this might seem restrictive, the study of Morse vector field on the sphere under the persistence setting is nontrivial and widely studied. 

\begin{figure}[!ht]
    \centering
    \includegraphics[width=0.90\columnwidth]{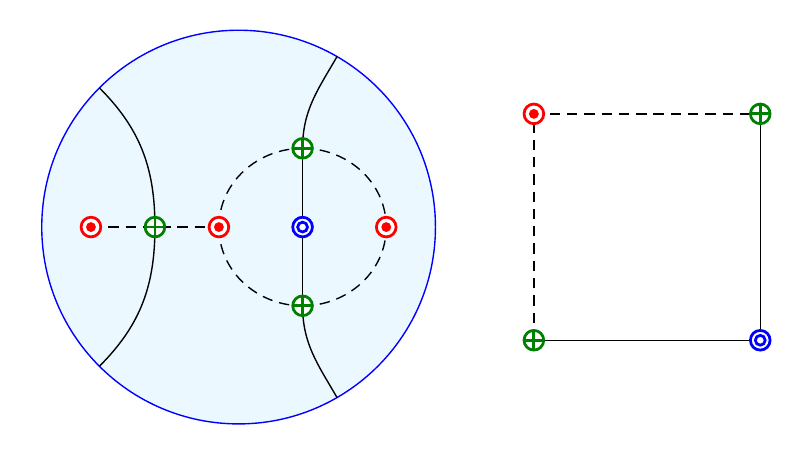}
    \vspace{-2mm}
    \caption{A topological skeleton of a Morse vector field on the sphere (left), and a quadrangle, or cell, of a Morse vector field (right).} 
    \label{fig:mvf}
\end{figure}

On a vector field, \emph{singularities} (or \emph{critical points}), including sources, saddles, and sinks, are locations where the vector values are zero. 
A \emph{streamline} is a line that is tangential to the instantaneous velocity direction. 
A \emph{topological skeleton} of a vector field consists of singularities and separatrices (streamlines connecting the singularities), which decomposes the domain into different modes of flow behavior.  
Figure~\ref{fig:mvf}(left) illustrates the topological skeleton of a vector field on the sphere, where the blue boundary indicates a (global) sink. Here, and further in this paper, \tikz{\maxx{0,0}} indicates a source, \tikz{\minn{0,0}} indicates a sink, and \tikz{\sadd{0,0}} indicates a saddle. 
Separatrices are either dashed lines (saddle-source connections) or solid lines (saddle-sink connections).

\para{Morse--Smale and Morse vector fields.}
Let $\TM_p$ denote the tangent space of $\Mspace$ at $p$.
A vector field $v$ on $\Mspace$ associates a vector $v(p) \in \TM_p$ to each point $p \in \Mspace$. 
An \emph{integral curve} of $v$ through a point $p \in \Mspace$ is a smooth map $\gamma\colon I \to \Mspace$ such that $\gamma(0) = p$ and $\gamma'(t) = v(\gamma(t))$ for all $t \in I$.
The image of an integral curve is called a \emph{trajectory}~\cite[page 10]{PalisMelo1982}. 
Two vector fields $v_1 \colon \Mspace_1 \to \Rspace^2$ and $v_2 \colon \Mspace_2 \to \Rspace^2$ are considered \emph{topologically trajectory equivalent} if there is a homeomorphism $h \colon \Mspace_1 \to \Mspace_2$ that transforms the trajectories of the vector field $v_1$ into the trajectories of the vector field $v_2$, preserving the orientations of the trajectories~\cite[Definition 1.1]{OshemkovSharko1998}.
A vector field $v$ is \emph{structurally stable} if the topological behavior of its trajectories is preserved under small perturbations of $v$ (that is, if the perturbed and the initial vector fields are trajectory topologically equivalent)~\cite[Definition 1.2]{OshemkovSharko1998}.

Suppose $\Mspace$ is compact. Then there exists a \emph{global flow} $\phi \colon \Rspace \times \Mspace \to \Mspace$ determined by $v$ such that $\phi(0,p) = p$ and $ \phi'(t,p) = v(\phi(x,p))$~\cite[Proposition 1.3]{PalisMelo1982}.   
For each $t \in \Rspace$, the map $v_t \colon \Mspace \to \Mspace$ is defined as  $v_t(p) = \phi(t, p)$. 
The \emph{$\omega$-limit set} of a point $p \in \Mspace$ is $\omega(p) =  \{q \in \Mspace \mid v_{t_n}(p) \to q \mbox{ for some sequence } t_n \to \infty\}$.  
The \emph{$\alpha$-limit set} of $p$ is $\alpha(p) = \{q \in \Mspace \mid X_{t_n} \to q \mbox{ for some sequence } t_n \to -\infty\}.$

A vector field $v$ on a closed two-dimensional surface is called a \emph{Morse--Smale vector field}~\cite[Definition 1.4]{OshemkovSharko1998} if
\begin{itemize}
\item $v$ has finitely many singular points and periodic trajectories, which are all hyperbolic;
\item There are no trajectories from a saddle to a saddle;
\item The $\alpha$-limit set and the $\omega$-limit set of each trajectory of $v$ is either a singular point or a periodic trajectory (a limit cycle).
\end{itemize}
A vector field is a \emph{Morse vector field} if it is a Morse--Smale vector field without periodic trajectories. 

A vector field is \emph{gradient-like for $f$} if it is topologically trajectory equivalent to the gradient vector field $\grad f$ of a function $f$ and a Riemannian metric $g_{ij}$ on $\Mspace$. 
Morse vector fields are precisely the gradient-like vector fields without saddle-saddle connections (separatrices from a saddle to a saddle)~\cite{Smale1961}. 
Finally, by~\cite[Quadrangle Lemma]{EdelsbrunnerHarerZomorodian2003}, every Morse vector field can be decomposed into regions, referred to as \emph{quadrangles} or \emph{cells}, as illustrated in Figure~\ref{fig:mvf}(right). 
In non-generic or boundary setting, a quadrangle may become degenerate, that is, a separatrix may have the same quadrangle on both of its sides.

\para{Morse functions.} 
A smooth function $f \colon \Mspace \to \Rspace$ defined on a manifold $\Mspace$ is a \emph{Morse function} if all critical points are non-degenerate. 
A Morse function is considered nice or well-behaved, as its critical points are isolated and stable (with respect to small perturbations).  
It plays an essential role in helping us understand the topology of a manifold.  

\para{Morse--Smale complexes.}
Let $f$ be a Morse function on a surface $\Mspace$ and $\grad{f}$ be its gradient. 
The \emph{stable manifold} $S(p)$ of a critical point $p$ of $f$ is the point itself together with all regular points whose integral lines end at $p$.
The \emph{unstable manifold} $U(p)$ of $p$ is is the point itself together with all regular points whose integral lines originate at $p$~\cite[Chapter 2]{EdelsbrunnerHarer2010}.  
A \emph{Morse--Smale function} is a Morse function whose stable and unstable manifolds intersect transversally. 
In this paper, we work in the restricted setting of Morse vector fields that arise as gradient vector fields of Morse--Smale functions, noting that we may work in the slightly more
general setting of gradient-like vector fields,
for which there is essentially no 
difference after modification of the metric $g_{ij}$~\cite[Theorem B]{Smale1961}.

For a given Morse--Smale function $f$, by intersecting the stable and unstable manifolds, we obtain the \emph{Morse--Smale cells} as the connected components of the set $U(p) \cap S(q)$ for all critical points $p, q \in \Mspace$~\cite{EdelsbrunnerHarerZomorodian2003}.  The \emph{Morse--Smale complex} (MSC) is the collection of Morse--Smale cells~\cite{EdelsbrunnerHarerZomorodian2003}. 
We define the \emph{Morse--Smale graph} of $f$ to be the $1$-skeleton of the MSC, that is, the union of the $0$-dimensional (vertices) and $1$-dimensional (edges) cells of the MSC of $f$; this is the \emph{topological skeleton}~\cite{HelmanHesselink1989} described earlier. 
In other words, the topological skeleton of the gradient vector flow of $f$ is equivalent to the 1-skeleton of the MSC of $f$.

\para{Persistence and persistence simplification.}
Persistent homology is a powerful tool in topological data analysis that is applicable in both data \emph{summarization} and \emph{simplification}. 
In its standard setting, persistent homology can be considered as an extension of Morse theory, in a sense that it studies homology groups of sublevel sets connected by inclusions, $\Mspace_a \xhookrightarrow{} \Mspace_b$ for $a \leq b$.  
In other words, it computes and summarizes topological features of a space  across multiple scales. 
The importance of a feature can be quantified via the notion of \emph{persistence}, that is, the amount of change to $f$ necessary to eliminate it~\cite{EdelsbrunnerMorozovPascucci2006}.  
Persistence is also useful in simplifying a function $f$ in terms of removing topological noise as determined by its persistence diagram or barcode~\cite{EdelsbrunnerLetscherZomorodian2002,EdelsbrunnerMorozovPascucci2006}. 

From now on, we consider only Morse vector fields and refer to them as vector fields throughout the remainder of this paper. 
Given (the topological skeleton of) a vector field, we construct a hierarchy by successive simplification based on persistence. 
Each step in the process cancels a pair of (adjacent) singularities (that is, indices contiguous with respect to a separatrix) and the sequence of cancellations is determined by the persistence of the pairs. 

\begin{figure}[!h]
    \centering
    \includegraphics[width=.99\columnwidth]{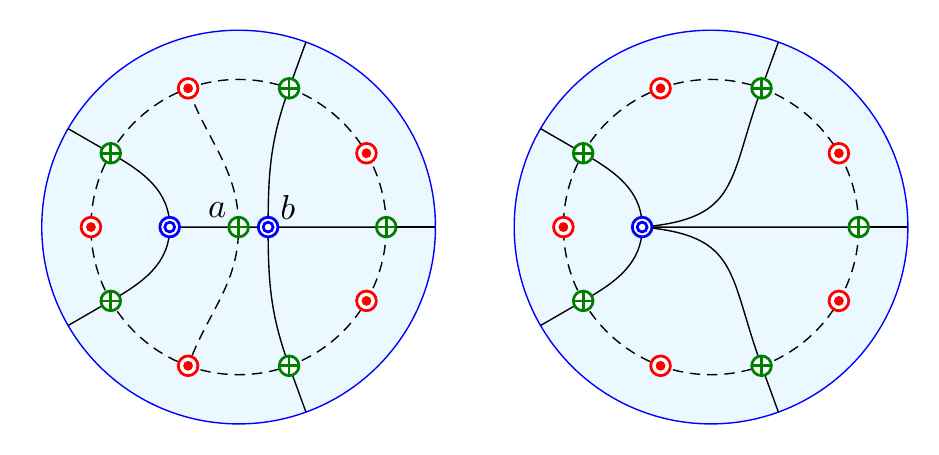}
    \vspace{-2mm}
    \caption{The topological skeleton before (left) and after the cancellation (simplification) of $a$ and $b$.
    } \label{fig:simplify}
\end{figure}

Heuristically, saddles cancel with sources or  sinks~\cite{EdelsbrunnerHarerZomorodian2003}.
We compute the pairing of singularities and their persistence using algorithms in~\cite{EdelsbrunnerHarerZomorodian2003}. 
The sequence of cancellations is in the order of increasing persistence, and the process can be simulated combinatorially.
Figure~\ref{fig:simplify} illustrates the cancelation of a sink-saddle pair. 
Note that for every positive $i$, the $i$-th pair of singularities ordered by persistence forms an arc in the topological skeleton obtained by canceling the first $i-1$ pairs~\cite[Adjacency Lemma]{EdelsbrunnerHarerZomorodian2003}. 
This means that, in general, not all singularities paired by the persistence algorithm are adjacent, however, they will become adjacent at the required time~\cite{EdelsbrunnerHarerZomorodian2003}. 
This lays the foundation for the persistence simplification of vector fields in our interactive system.

\section{Related Work}
\label{sec:related-work}

\para{Vector field design systems.}
A few vector field design systems have been created for domain-specific  applications, such as graphics~\cite{PraunFinkelsteinHoppe2000, Turk2001, WeiLevoy2001, Hertzmann1998, HertzmannZorin2000} and fluid simulation~\cite{Stam2003}. 
The techniques could be roughly classified into projection, diffusion, and interpolation.
In the first approach, a 3D vector field is specified and projected onto the surface~\cite{Wijk2003}. This is fast and simple, however difficult to achieve fine-grained control. 
In the second approach, the system performs relaxation based on a set of user-specified vectors, where known vector values are propagated like diffusion  across the remaining surface~\cite{Turk2001, WeiLevoy2001}. 
In the third approach, a global vector field is built by interpolating a few user-specified vector values using basis functions~\cite{PraunFinkelsteinHoppe2000, Wijk2002}.  

To provide more control over vector field topology, other design systems focus on specifying the number, types and locations of singularities or to a larger extent, the topological skeleton. 
A simple way to design and control a vector field begins with a set of \emph{flow primitives} or \emph{building blocks}~\cite{WejchertHaumann1991}, and such primitives are combined into a global vector field. 
For instance, a primitive can be a simple vector field in the local neighborhood of a user-specified singularity; and multiple primitives can be combined using radial basis functions~\cite{Wijk2002}.  
Furthermore, singularities can be added, removed or edited by users~\cite{RockwoodBinderwala2001}. 
The users can also specify the entire topological skeleton of the desired vector field~\cite{Theisel2002}, though a complete specification may be inefficient.

In terms of vector field simplification, techniques are often based on Laplacian smoothing~\cite{PolthierPreuss2003,WestermannJohnsonErtl2000,TongLombeydaHirani2003}, while topology-based techniques originate from the study of Morse theory and gradient vector fields of Morse functions. 
By specifying the number and configuration of the critical points of a Morse function and running multi-grid relaxation, the design of Morse functions over a surface is equivalent to the design of their gradient vector fields~\cite{NiGarlandHart2004}. 
The gradient vector field of a scalar function can be simplified by modifying function values near the singularity pair~\cite{EdelsbrunnerLetscherZomorodian2002,EdelsbrunnerHarerZomorodian2003}. 
A singularity pair can also be simplified directly by performing nonlinear optimization surrounding the pair~\cite{TricocheScheuermannHagen2001} or using Conley index theory~\cite{ZhangMischaikowTurk2006}. 

Our system is topology-based and shares some similarities with the above systems, with three main distinctions:
\begin{enumerate}
\item Elementary moves (Section~\ref{sec:methods}) are used as fundamental building blocks in designing the vector fields incrementally;
\item Much finer control is given to the vector field topology, in particular, adjusting the geometry of separatrices. 
\item Persistence barcodes are used explicitly to guide interactive vector field simplification and exploration. 
\end{enumerate}

\para{Topological classification of Morse(--Smale) vector fields.}
Our work is related to the topological classification of Morse(--Smale) vector fields with one crucial difference: We are interested in the equivalence classes of Morse functions (and their gradient vector fields) that share the same persistence barcode.  

One of the first invariants defined for Morse--Smale vector fields on a closed surface is Peixoto's \emph{distinguished graph}, that is, a graph together with a distinguished set of edges satisfying some conditions~\cite{Peixoto1973}.
The distinguished graph provides an invariant for Morse--Smale vector fields on
a surface, but has a very technical description. 
The graph corresponds to connected components of what is left after deleting saddle singularities and their associated stable and unstable manifolds from the surface. 
Several simpler invariants have been proposed based on Peixoto's work.
\emph{Cyclic distributions
of colored points} introduced by Fleitas is a simplification of the distinguished graph invariant~\cite{Fleitas1975}, and the
 \emph{coloured dual graph} by Wang is a variant for orientable closed surfaces~\cite{Wang1990}.
All three mentioned above are invariants of Morse--Smale vector fields on closed surfaces, and they become complete invariants when restricted to Morse vector fields (see~\cite{OshemkovSharko1998} for an extended comparison of these
invariants). 

Related to classification of gradient-like vector fields on a surface is the classification of functions themselves. 
An \emph{$a$-function} is a Morse function on a surface with exactly three critical values.
The \emph{$f$-invariant} was constructed to classify 
$a$-functions up to conjugacy~\cite{oshemkov1994morse}, and hence their vector fields up to topological equivalence\cite[Remark 2.8]{oshemkov1994morse}.
In addition, there has been work classifying Morse
functions on surfaces, albeit from a much different perspective~\cite{Nicolaescu2008,Arnold2007,Sharko2003}.

Some of the mentioned invariants could also be used in the design of vector fields, left open as an avenue for future research, however, they do not lead to as simple and efficient operations as the elementary moves (Section~\ref{sec:methods}) employed in this paper.

\section{Methods}
\label{sec:methods}

We describe the main analytic components within {\tool}, including elementary moves as fundamental building blocks, vector field construction with basis functions, and geometric control of separatrices using splines. 

\subsection{Elementary Moves}

\begin{figure}[!h]
    \centering
    \includegraphics[width=.95\columnwidth]{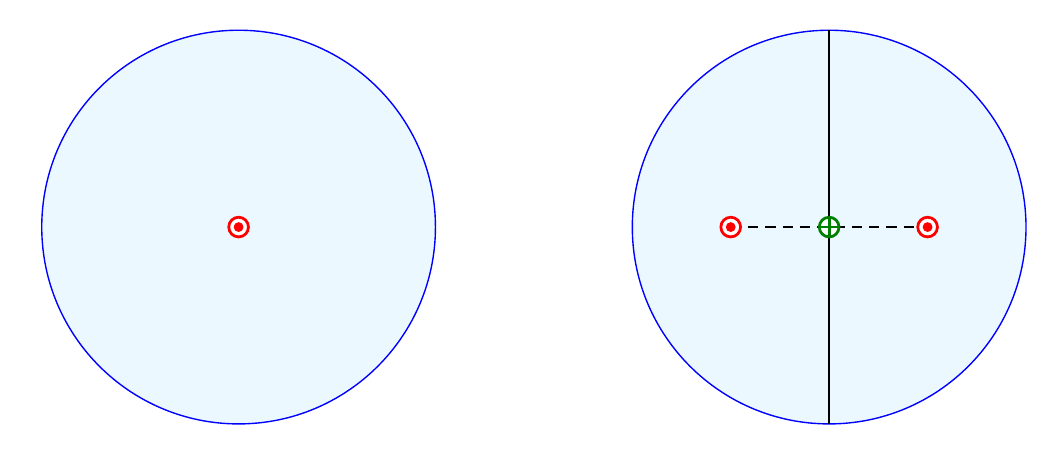}
    \vspace{-2mm}
    \caption{The simplest vector field on a sphere (left) and the simplest Morse--Smale vector field on a sphere (right).
    } \label{fig:default}
\end{figure}

Since the Euler characteristic of the sphere is 2, the simplest vector field on a sphere is one with a single source and a single sink, see Figure~\ref{fig:default}(left). 
We visualize the sink as a boundary cycle in blue: imagine ``flattening" the sphere onto a disk by expanding a rubber band surrounding the sink. 
However, a Morse--Smale vector field on the sphere must contain at least one saddle~\cite[Quadrangle Lemma]{EdelsbrunnerHarerZomorodian2003}, which must have four edges coming out of it, which can not be identified with each other. Such a configuration is realizable on the sphere, as in Figure~\ref{fig:default}(right), which we use as the \emph{default configuration} for {\tool}.

\begin{figure}[!h]
    \centering
    \includegraphics[width=.99\columnwidth]{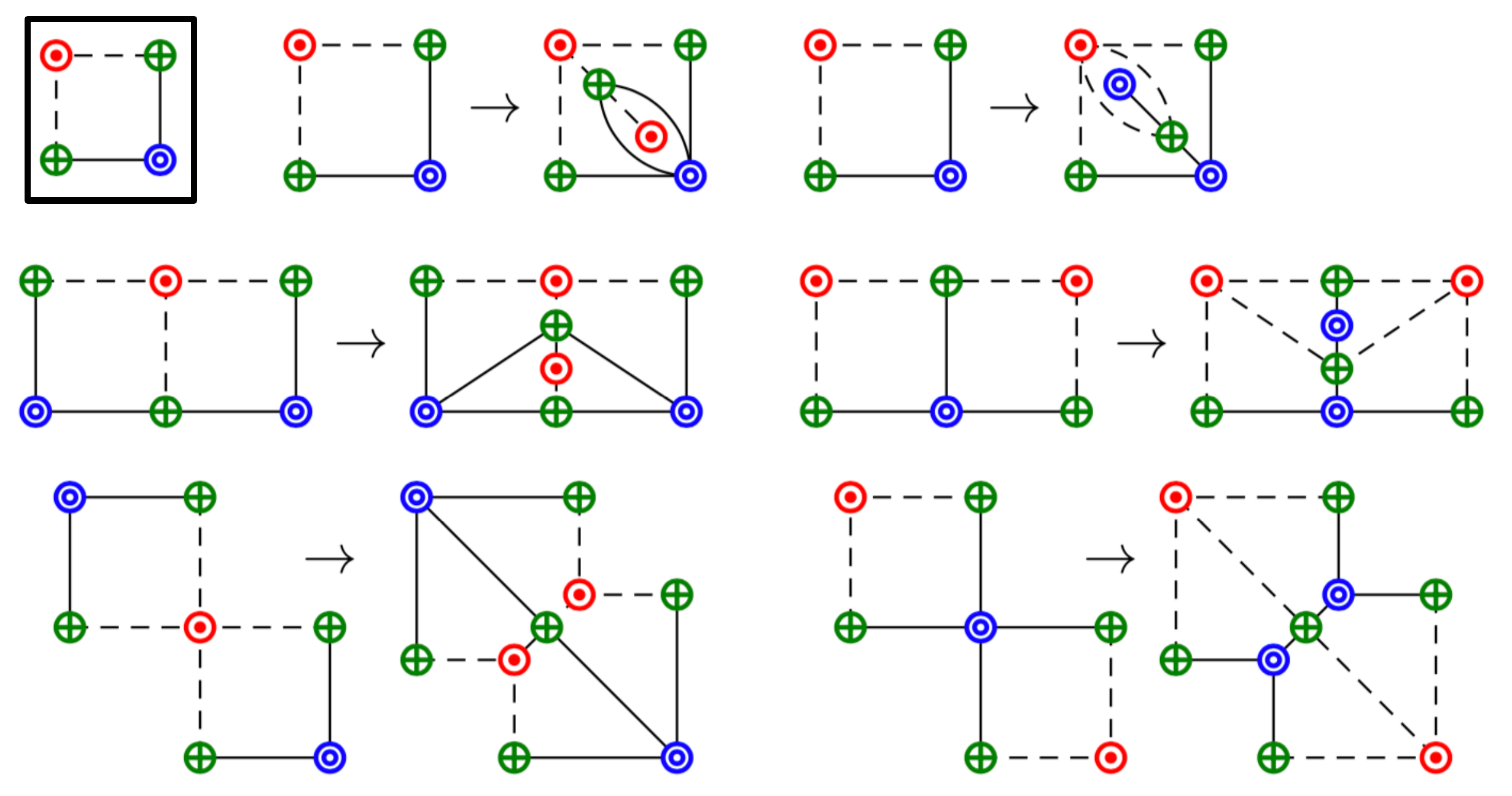}
    \vspace{-2mm}
    \caption{Elementary moves and a cell from a Morse vector field (top left). The elementary moves (left to right) comprise face-max and face-min moves (top row), edge-max and edge-min moves (middle row), vertex-max and vertex-min moves (bottom row). 
    } \label{fig:moves}
\end{figure}

Our visualization framework is based upon understanding how cells, generically as quadrangles, in Figure~\ref{fig:moves}(top left), of a vector field can fit together on a surface and how they change when a pair of singularities is added to their interiors and boundaries; such operations are referred to as \emph{elementary moves}. 
An elementary move is an action that initiates or advances our design process. 
Elementary moves originate from a mathematical  framework~\cite{CatanzaroCurryFasy2019} that studies different notions of equivalence for Morse functions on the sphere in the context of persistent homology. 

The following theorem is a corollary of~\cite[Theorem 3]{CatanzaroCurryFasy2019} that defines elementary moves on a vector field, with the goal of describing all possible ways to create a new vector field. 

\begin{theorem}[Elementary Move Theorem].
Any two Morse vector fields are related to each other through a sequence of face, edge, and vertex moves.
\end{theorem}

These moves are illustrated in Figure~\ref{fig:moves}. By the Quadrangle Lemma~\cite{EdelsbrunnerHarerZomorodian2003}, every cell of a Morse vector field is bounded by four edges (counting an edge twice if the cell is on both sides of the edge).
This allows us to describe changes to the vector field as a composition of moves.
Everything in the vector field outside of this region stays the same between moves.
Specifically, a \emph{face move} adds one saddle-source or saddle-sink pair in the interior of a cell.
An \emph{edge move} adds one saddle-source or saddle-sink pair on the edge of a cell. 
A \emph{vertex move} adds one saddle-source or saddle-sink pair at an existing singularity.
All moves add two cells to the vector fields.  
Notice that all moves are, in fact, reversible; that is, the addition of a pair of singularities translates to the removal of a pair of singularities in the reverse direction. 
The face, edge, and vertex moves are ways of manipulating the vector field to obtain another vector field. 
These moves by themselves do not have functional values associated with the singularities. 

However, since the gradient of a Morse(--Smale) function gives rise to a Morse vector field, if we combine the modification of the gradient vector field with the modification of function values at the singularities (see Section~\ref{sec:interface}), we could equivalently construct and explore the space of Morse functions.   

\subsection{Initialization, Control, Debugging, and Simplification}

\para{Initial vector field design.}
Once the user specifies the types and locations of singularities via the elementary moves, we use the framework of Zhang et al.~\cite{ZhangMischaikowTurk2006} to construct an initial vector field. The vector field is represented as a triangulated mesh with vector values assigned at the vertices of the mesh.
We attach a \emph{basis vector field} to each (user-specified) singularity, and construct a designer vector field as the truncated sum of these basis vector fields.
For instance, a basis vector field centered at a source $\textbf{p}_0 = (x_0, y_0)$ is defined as:
\[
V(\textbf{p})=e^{-d\|\textbf{p} - \textbf{p}_0 \|^2}
\begin{pmatrix} 
k & 0 \\
0 & k 
\end{pmatrix}
\begin{pmatrix} 
x-x_0 \\
y-y_0 
\end{pmatrix},
\]
for any point $\textbf{p} = (x,y) \in \Rspace^2$, where $d$ is a constant that is used to control the amount of influence of the basis vector field, and the matrix $\begin{pmatrix} 
k & 0 \\
0 & k 
\end{pmatrix}$ indicates the type of the singularity. 
For sinks and saddles, we replace this matrix with 
$\begin{pmatrix} 
-k & 0 \\
0 & -k
\end{pmatrix}$ and $\begin{pmatrix} 
k & 0 \\
0 & -k 
\end{pmatrix}$, respectively.

\para{Geometric control of the separatrices.} 
To provide a high level of control of the separatrices, we approximate their geometry using cubic cardinal splines with tension 0. In addition to the initial vector field, we also generate another auxiliary vector field which captures the flow along separatrices. The final vector field is a weighted sum of the initial and the auxiliary vector field, where the weight is computed from the distance between a mesh vertex and its closest separatrix. We apply additional smoothing in the neighborhood of separatrices to prevent the creation of spurious singularities, by replacing the function value of each mesh vertex in the neighborhood of separatrices with the weighted average function value of its neighbors. The weights are inversely proportional to the distances between the vertex and its neighbors.

\para{Debugger.} 
{\tool} provides great flexibility for a user to control geometric details involving separatrices. 
A key component for vector field construction is the \emph{debugger}, which detects invalid configurations. 
Invalid configurations may arise due to user operations, semi-automatic modes, simplification, or boundary conditions. Flow animation is not allowed when the debugger detects a configuration to be invalid.   

In this paper, we assume all saddles are simple and all higher-order saddles can be unfolded into simple saddles. 
Every saddle therefore is of degree four, and the endpoints of its four adjacent separatrices alternate between connections with sources and sinks, as in Figure~\ref{fig:mvf}, for example.  
Sources and sinks may have arbitrary degrees. 
A debugger will report an invalid configuration if:
\begin{itemize}
\item The separatrices adjacent to a saddle do not follow the appropriate \emph{alternating} order. Recall a saddle-source connection is indicated by a solid line, and a saddle-sink connection is marked by a dashed line. The configuration of a saddle is invalid if its adjacent separatrices are not in alternating solid and dashed lines. 
\item The end point of a separatrix is not properly attached to a singularity or the boundary (the global sink).    
\item The separatrices are crossing.
\item A singularity is isolated without any separatrix attachment, with the exception of the minimal configuration in Figure~\ref{fig:default}(left). 
\item There is a saddle-saddle separatrix. 
\item A singularity is dragged outside the boundary.
\end{itemize}
See Section~\ref{sec:interface} for examples of invalid configurations detected by the debugger. 

\para{Vector field simplification.}
We use Perseus~\cite{MischaikowNanda2013} for efficient computation of persistent homology, which gives rise to persistence pairings of singularities. 
Since some of these pairs are in fact adjacent, based on~\cite[Adjacency Lemma]{EdelsbrunnerHarerZomorodian2003} (Section~\ref{sec:background}), they can be simplified by modifying the basis functions defining these singularities. 
Each simplification operation removes a pair of adjacent singularities ranked by persistence, together with edges adjacent to the pair; resulting in a simplified topological skeleton. 
See Figure~\ref{fig:simplify} for an example of simplifying a saddle-sink pair $(a,b)$ connected by an edge. 

\subsection{System Design}
{\tool} is web-based and can be accessed from any modern web-browser (tested using Google Chrome and Mozilla Firefox). {\tool} is implemented in HTML, CSS, and JavaScript, with Python and Flask as the backend server to handle requests from the browser. The module collection \texttt{D3.js} is used for rendering SVGs, and the flow animation is generated with a Canvas element. Perseus~\cite{MischaikowNanda2013} is used to compute persistence homology and produce barcode. {\tool} is provided open-sourced via GitHub\footnote{https://github.com/zhou325/VIS-MSVF}.
We also provide a supplementary video that demonstrates the capabilities of {\tool}.

\section{The {\tool} User Interface}
\label{sec:interface}

The user interface of our design system is provided in Figure~\ref{fig:interface}, see the supplementary video for a demo. The system contains five main interactive panels: the flow visualization panel, the elementary moves panel,  the function adjustment panel, the history panel and the barcode panel.  

\begin{figure}[!ht]
    \centering
    \includegraphics[width=.98\columnwidth]{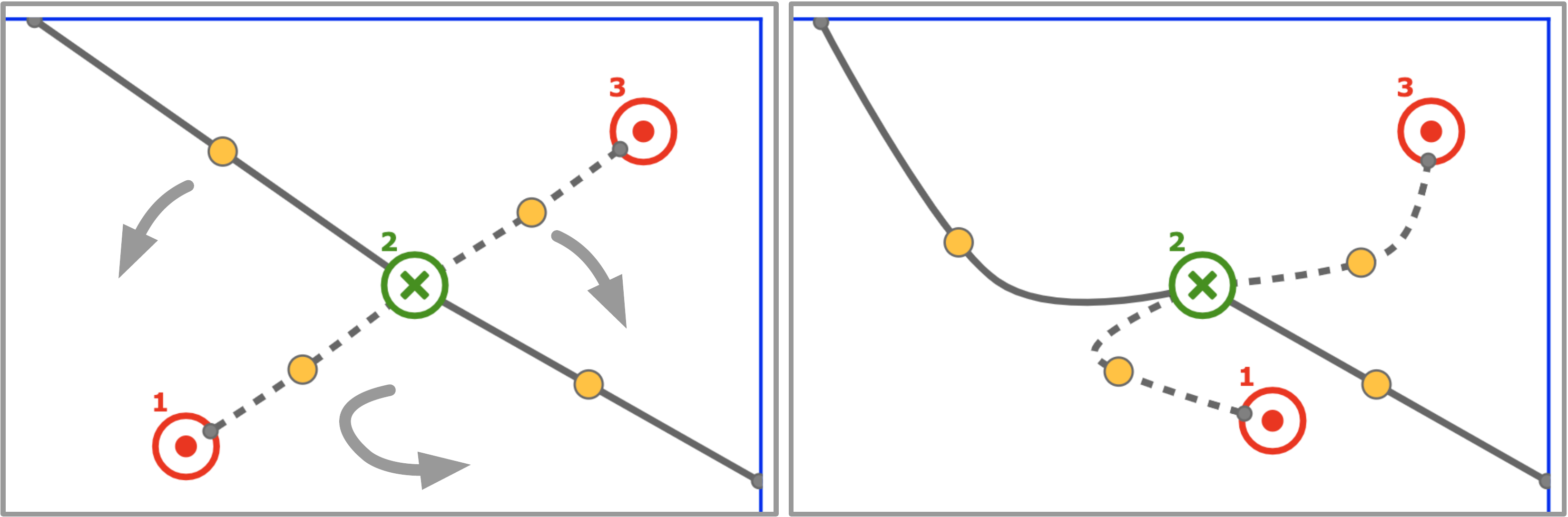}
    \vspace{-2mm}
    \caption{Control points (in yellow) are used to modify (left to right) the geometry of separatrices surrounding a saddle point.}
    \label{fig:spline}
    \vspace{-2mm}
\end{figure}

\para{Flow visualization panel.}
Flow visualization panel (A) supports modifying the topology and geometry of the topological skeleton. 
In particular, as illustrated in Figure~\ref{fig:spline}, users can modify the locations of singularities by ``drag and drop". 
Since separatrices are modeled as splines, users can also modify the geometry of separatrices using yellow control points of the splines. 

The panel supports both static and dynamic flow visualization. The topological skeleton, the (static) flow visualization, and the animation of the current configuration can be enabled or disabled as desired. 
The animation is particular useful for the user to get an intuitive sense of the dynamics of the designer vector field, as in Figure~\ref{fig:interface}(A).

\begin{figure}[!ht]
    \centering
    \includegraphics[width=.8\columnwidth]{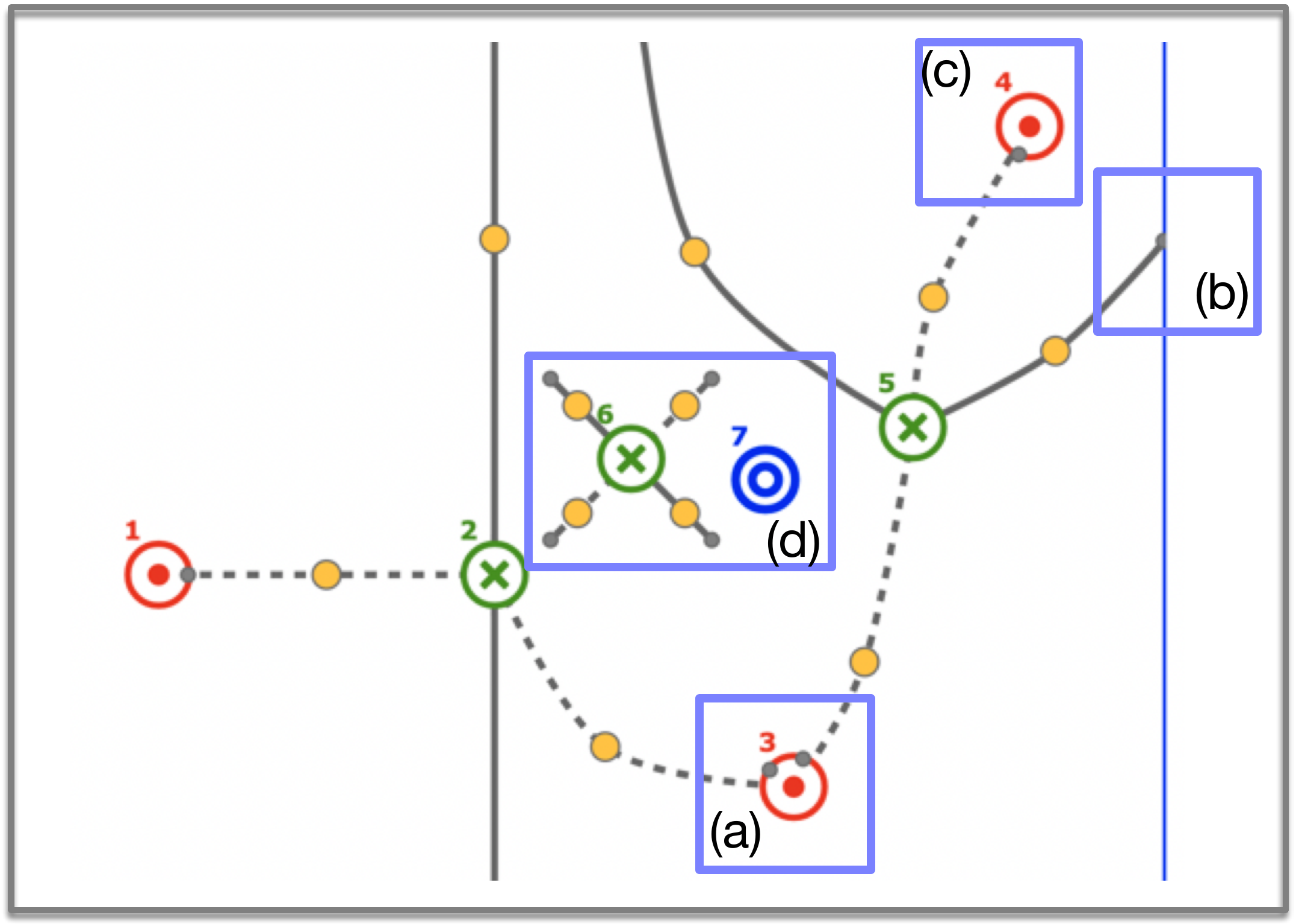}
    \vspace{-2mm}
    \caption{Various ways of connecting pairs of singularities.
    (a) A source is attached to two saddles, and the attachment map is defined by gluing the end points (in grey) of saddle-source connections.
    (b) A saddle is attached to the boundary (the global sink).
    (c) A source is attached to a saddle via a single attachment point.
    (d) The initialization of a face-min move under manual mode. }
    \label{fig:attach}
    \vspace{-2mm}
\end{figure}

\para{Elementary moves panel.}
Various elementary moves are provided via the elementary moves panel of Figure~\ref{fig:interface}(B).  
Under \textit{manual mode}, a user connects pairs of singularities manually and our system checks for valid configurations. 
Using \textit{semi-automatic mode}, separatrices are added automatically, followed by user adjustments. 
Figure~\ref{fig:attach} shows various ways of connecting pairs of singularities.
{\tool} provides fine-grained control in attaching edges to singularities.
Specifically, the end points (in grey) of each separatrix can be attached to the circular boundaries of the glyphs representing different types of singularities. 

\begin{figure}[!ht]
    \centering
    \includegraphics[width=.7\columnwidth]{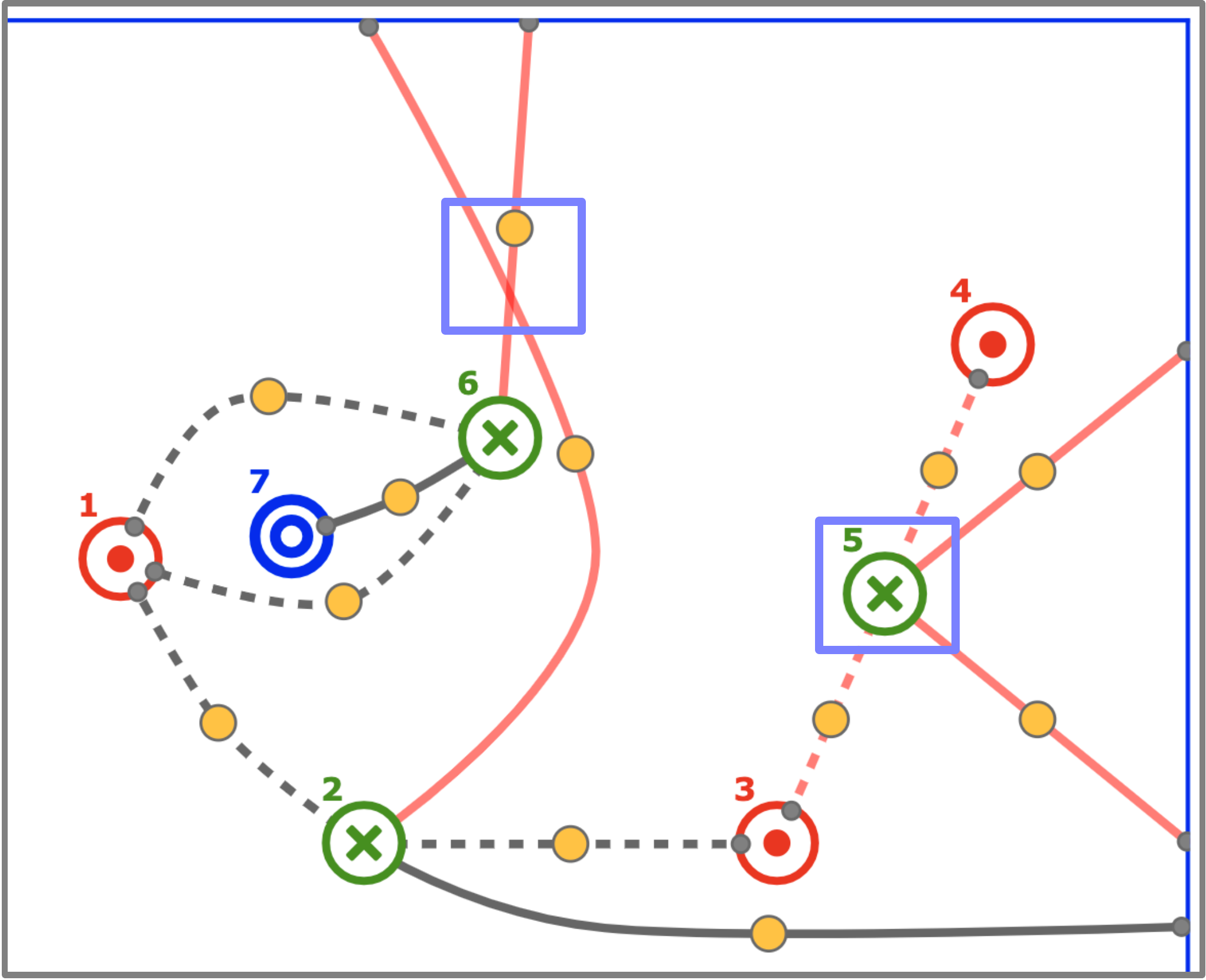}
    \vspace{-2mm}
    \caption{Examples of invalid configuration, where invalid seperatrices are highlighted in red.}
    \label{fig:checker}
\end{figure}

For both manual and semi-automatic mode, a key component connecting the flow visualization panel with the elementary moves panel is the \emph{debugger} (Section~\ref{sec:methods}).
It detects invalid configurations when a user is modifying the geometry of the topological skeleton during the design process, or when the semi-automatic configuration of an elementary move does not resolve all errors. 
When an invalid configuration is detected, edges involved in the configuration are highlighted in red; additional warning messages are provided to the user to guide necessary correction and adjustment operations, see Figure~\ref{fig:checker} for an example. 
There, blue boxes enclose regions of error, where either edges are intersecting, or the edges surrounding a saddle are not in alternating order.

\para{Function adjustment panel and history panel.} 
Recall that the design of a Morse function is equivalent to the design of their gradient vector fields, and a gradient vector field can be modified by modifying function values at the singularities. 
The function adjustment panel in Figure~\ref{fig:interface}(C) enables a user to modify the function values at singularities. 
Such modification does not necessarily change the flow directions, but may change the flow magnitude. 
{\tool} automatically checks for constraints in terms of function values during the adjustment operation, that is, it ensures the function value of saddle is bounded above by function values of its adjacent local maxima, and bounded below by its adjacent local minima. 
The history panel in Figure~\ref{fig:interface}(D) stores all valid and invalid operations during manipulations and supports redo and undo operations.  

\begin{figure}[!ht]
    \centering
    \includegraphics[width=.98\columnwidth]{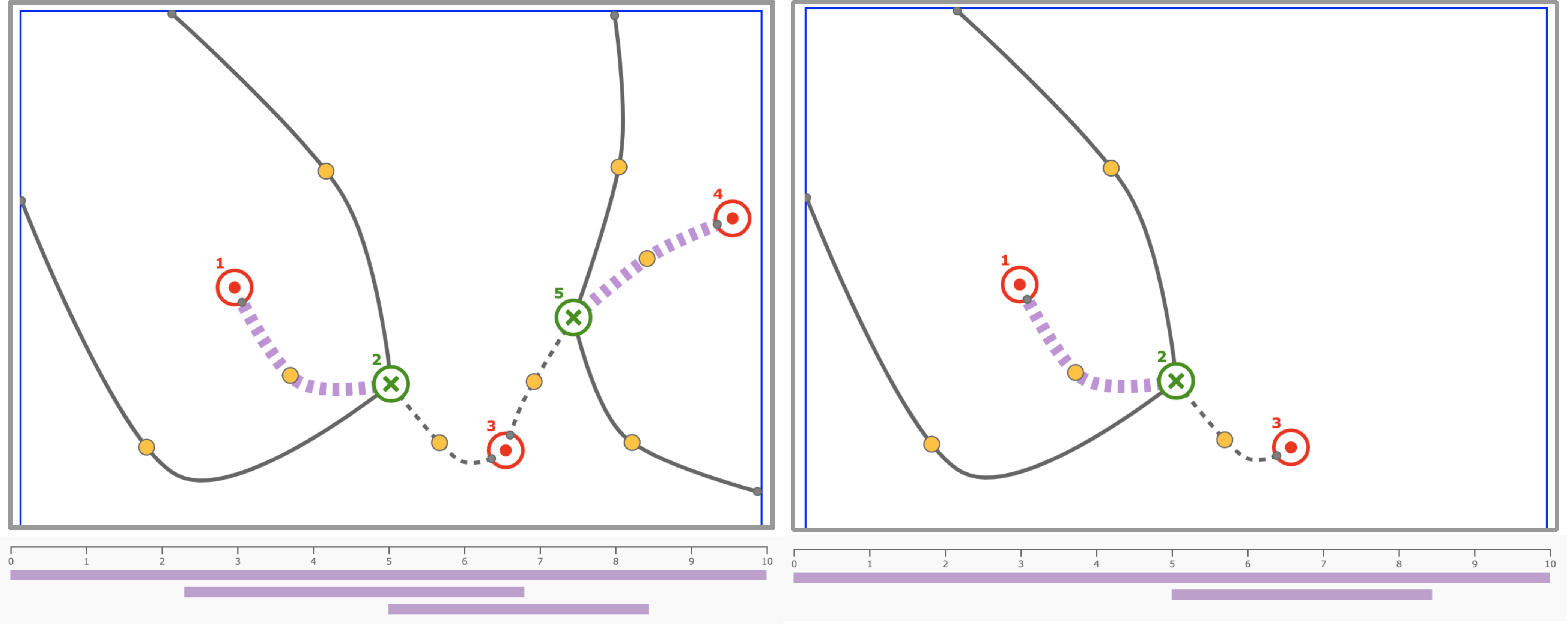}
    \vspace{-2mm}
    \caption{All simplification candidate pairs are connected by purple dotted lines (left), any of which may be chosen for simplification (right).}
    \label{fig:barcode}
\end{figure}

\para{Barcode panel.}
Finally, using the barcode panel in Figure~\ref{fig:interface}(E), we provide the persistence barcode of the current vector field configuration, as well as interactive capabilities for persistence-based vector field simplification.
As illustrated by a simple example in Figure~\ref{fig:barcode}, when the simplification button is clicked (left), the flow visualization panel highlights adjacent pairs of singulars that are eligible for simplification, marked by purple dotted lines. Selecting a particular bar leads to the highlighting of a potential candidate pair, if one exits, and clicking on an eligible pair will simplifying the underlying field (right). 
The remaining configuration contains a last candidate pair for simplification, and simplifying it will result in the configuration of Figure~\ref{fig:default}(left).






\section{Usage Scenarios}
\label{sec:results}

We describe various usage scenarios to illustrate the powerful capabilities of {\tool} for users beyond visualization and graphics researchers, in particular, mathematicians. 
{\tool} will help topologists, geometers, and combinatorialists explore invariants in the classification of flows and characterize Morse functions in the persistence setting, to name a few.  

\subsection{Studying Topological Invariants in Flow Classification}
\label{sec:topological-invariants}

Qualitative analysis of dynamical systems and the classification of such systems is an active area of research, which has wide-spread applications in mathematics, physics, biology, economics, and medicine. 
Users of {\tool} can study various topological invariants employed in the classification of flows on surfaces, as the elementary moves give insight into the construction of certain invariants and highlight the differences between them (Section~\ref{sec:related-work}). 

\begin{figure}[!h]
    \centering
    \includegraphics[width=.99\columnwidth]{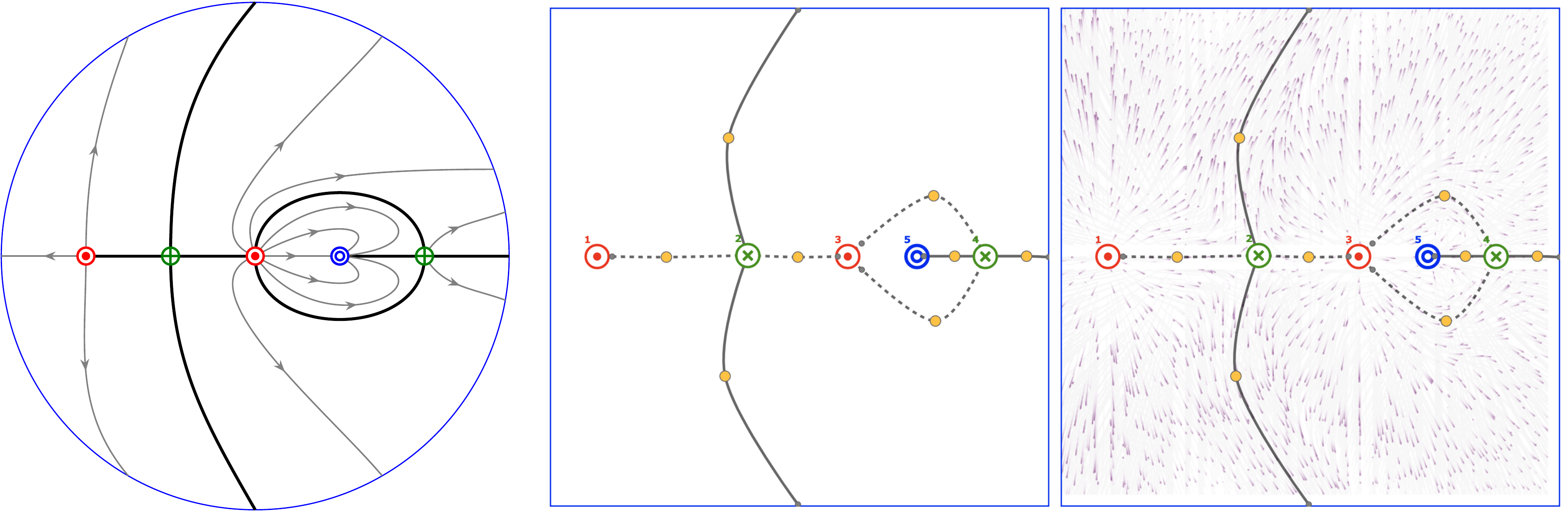}
    \vspace{-2mm}
    \caption{The design of a vector field on the sphere. An example (left) reproduced from~\cite[Example 1.8]{OshemkovSharko1998} used to study the three-color graph, a construction (middle) of the same example using {\tool}, and its flow visualization (right). 
    } \label{fig:flow-on-sphere}
\end{figure}

For example, Oshemkov and Sharko~\cite{OshemkovSharko1998} studied an invariant of a Morse flow called a \emph{three-color graph} and proved that such an invariant classifies Morse flows on two-dimensional surfaces up to trajectory topological equivalence. 
Example 1.8 from~\cite{OshemkovSharko1998}, reproduced in Figure~\ref{fig:flow-on-sphere}(left), is a flow defined on a 2-sphere containing two sources, two saddles and two sinks.
One of the sinks is on the reverse side of the sphere, which is represented by the blue boundary circle; separatrices are shown in bold.

Here we demonstrate that {\tool} can easily reconstruct interesting flows in the literature, such as the example in Figure~\ref{fig:flow-on-sphere}(left). 
In this case we have used a single elementary move, a face-min move, on the default configuration, which is a standard height function on the 2-sphere (a pair of sources, a boundary sink, and a saddle), to obtain a vector field with two sources, two sinks, and two saddles, see Figure~\ref{fig:flow-on-sphere}(middle). 

Other authors have constructed different invariants for Morse and Morse-Smale flows on (orientable) surfaces.
In addition to the three-color graph, Peixoto~\cite{Peixoto1973} introduced a \textit{distinguished graph}, Fleitas~\cite{Fleitas1975} introduced \textit{cyclic
distributions of coloured points}, and Wang~\cite{Wang1990} introduced a \textit{coloured
dual graph} - all of which can be described and their effects explored using {\tool}. We replicate Example 1.8 of~\cite{OshemkovSharko1998} since it is simple, yet still shows enough complexity of the general theory. 

\subsection{Characterizing Morse Functions Through Persistence}
\begin{figure}[!h]
    \centering
    \includegraphics[width=.9\columnwidth]{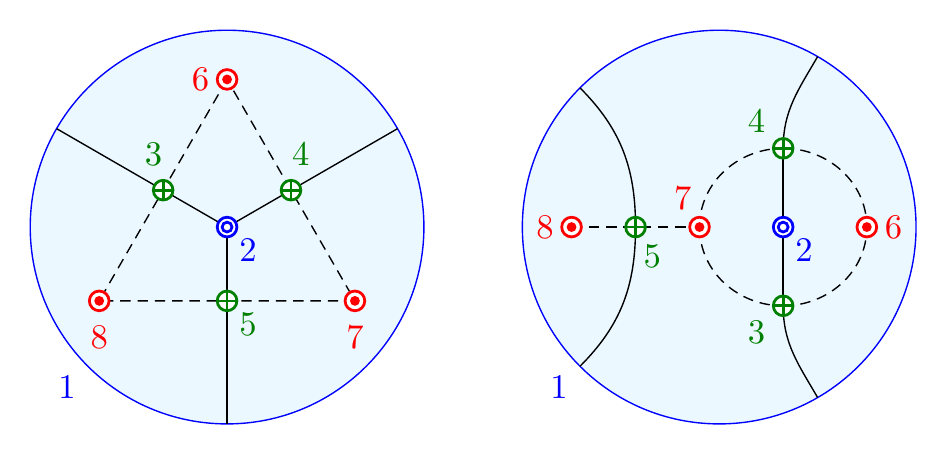}
    \vspace{-2mm}
    \caption{An example of two Morse functions on the sphere that have the same sub-level set barcode, with critical values indicated. These are \emph{not} graph equivalent functions.} 
    \label{fig:graph-equal}
\end{figure}

Our original motivation for developing {\tool} was inspired by a mathematical framework~\cite{CatanzaroCurryFasy2019} that investigates different moduli spaces of Morse functions from the perspective of persistence. 
In this vein, we are interested in using {\tool} to characterize the set of Morse functions that give rise to the same barcode.  
Two Morse functions may give rise to the same barcode, but they are not necessarily considered equivalent if taking one function to another requires a significant amount of deformation; see Figure~\ref{fig:graph-equal} for an example. 
Recall two Morse--Smale functions $f,g$ are \emph{graph equivalent} if there is a graph isomorphism between their Morse--Smale graphs. 
Consider the two Morse functions on the sphere in Figure~\ref{fig:graph-equal}, which have the same barcode. Note that these functions are are not graph equivalent. That is, a deformation from the function given in Figure~\ref{fig:graph-equal}(left) to the function given in Figure~\ref{fig:graph-equal}(right) requires significant perturbation. 

\begin{figure}[!h]
    \centering
    \includegraphics[width=.98\columnwidth]{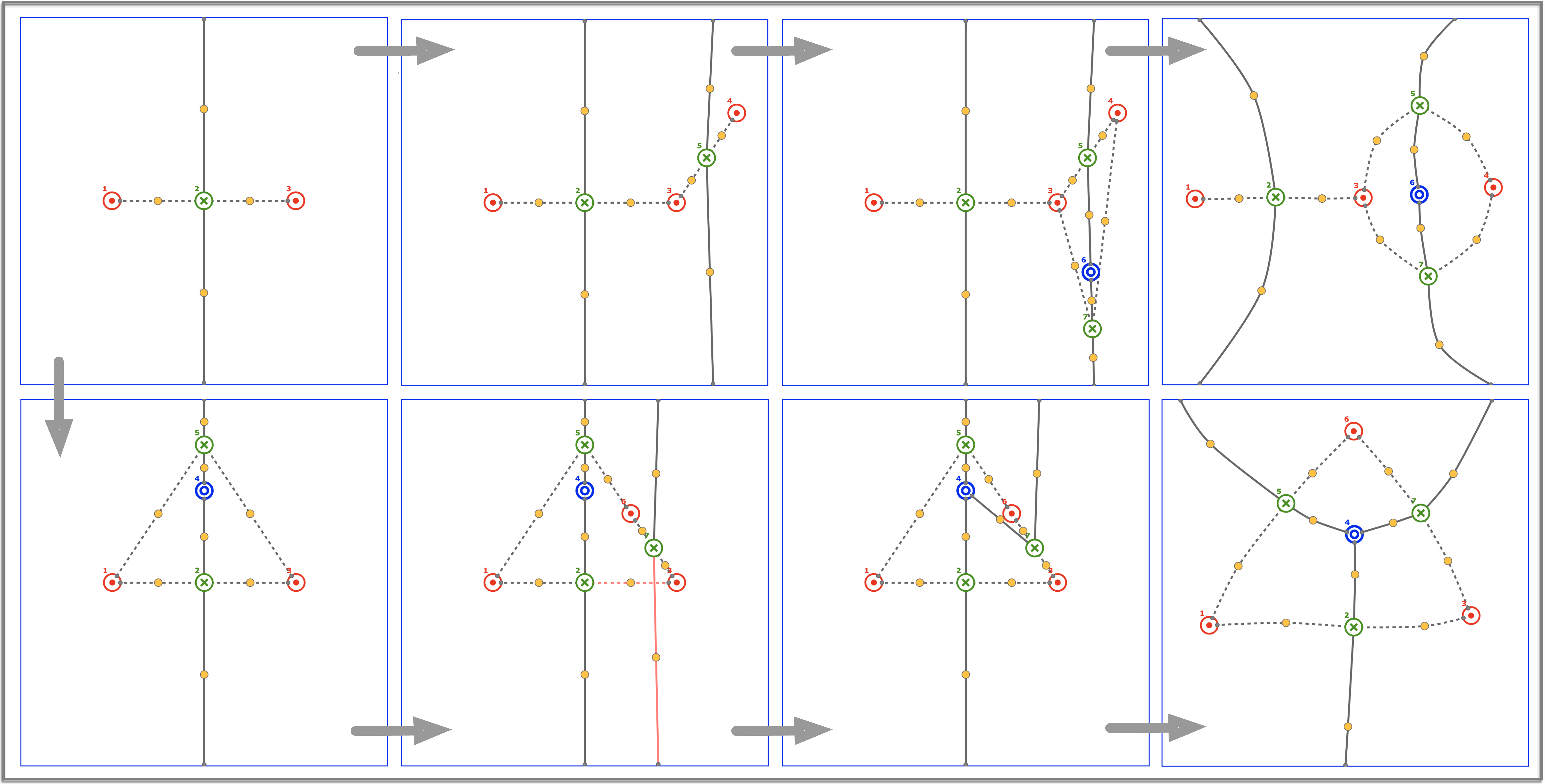}
    \vspace{-2mm}
    \caption{Using {\tool} to recreate Morse functions on the sphere that have the same barcode but are not graph equivalent.} 
    \label{fig:graph-equal-design}
\end{figure}

Using {\tool}, Figure~\ref{fig:graph-equal-design} shows we can easily recreate the two configurations.
Starting from the default configuration, the configuration of Figure~\ref{fig:graph-equal}(left) can be created under the semi-automatic mode, using a face-max and an edge-min move, followed by geometric modifications to the separatrices. 
 The configuration Figure~\ref{fig:graph-equal}(right) can be generated semi-automatically by an edge-min move and an edge-max move, in combination with geometric operations. 
 Notice that the semi-automatic edge-max move creates an invalid configuration (detected by the debugger), which is subsequently corrected.  

Hence through~\cite{CatanzaroCurryFasy2019} we can explore the space of Morse functions (that give rise to the same barcode) by putting different equivalence relations on the space. 
Each choice of equivalence relation leads to a different moduli space structure on the space of Morse functions, and each equivalence class has an interesting combinatorial structure that can be used practically to enrich the barcode. 
Given {\tool}, we can start to address the following question: 
Can we characterize equivalence classes of Morse functions on the sphere that have the same barcode?

\subsection{Designing and Visualizing Morse--Smale Complexes}

Given the connection between Morse functions, Morse vector fields and MSCs, {\tool} can also be utilized to design MSCs. 
The design process offers insights into structural variations of MSCs in terms of persistence simplification, which is useful in constructing MSC ensembles for uncertainty quantification and uncertainty visualization. 

\begin{figure}[!h]
    \centering
    \includegraphics[width=.98\columnwidth]{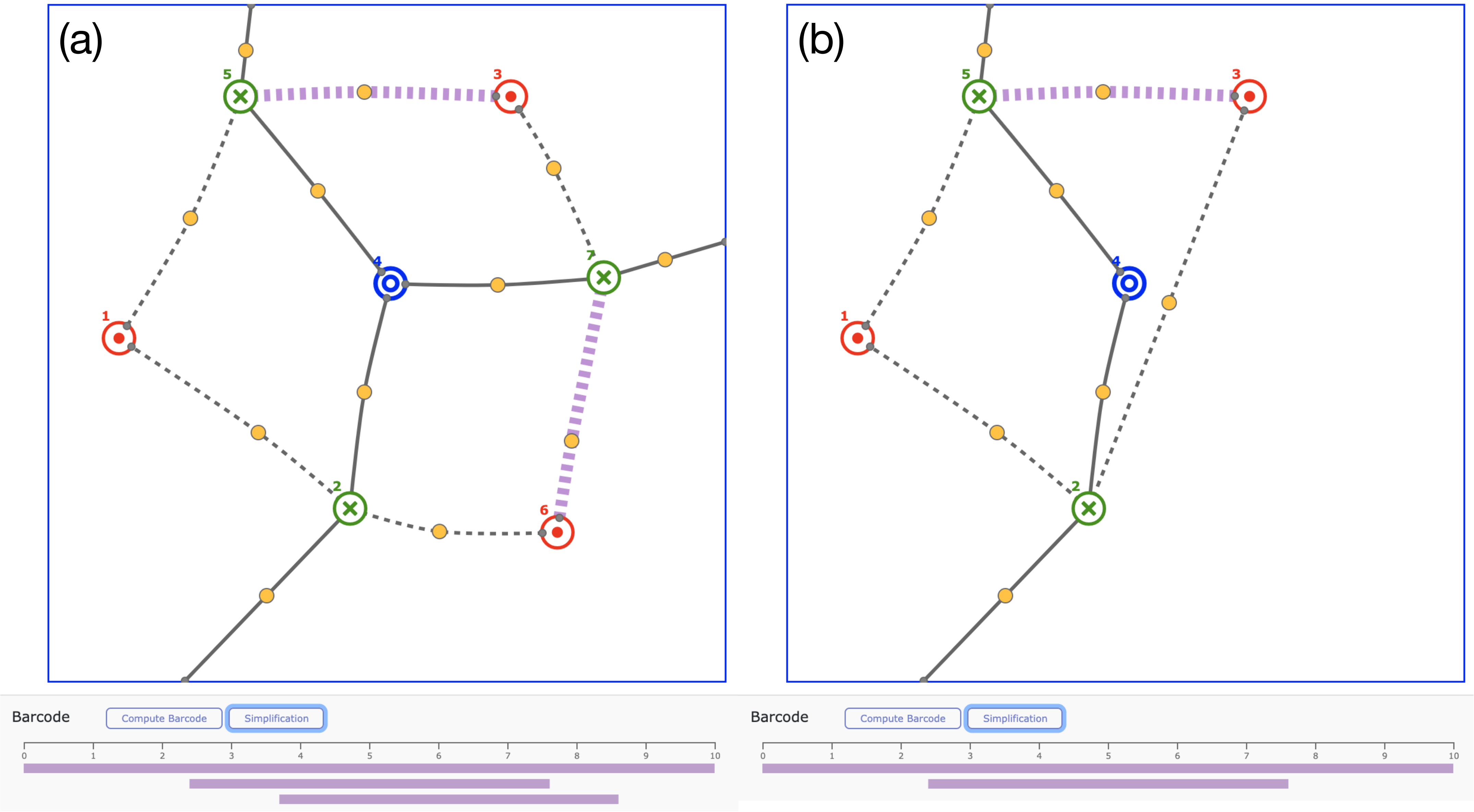}
    \vspace{-2mm}
    \caption{Using {\tool} to design Morse--Smale complexes. A Morse--Smale complex is shown together with its barcodes before (a) and after (b) simplification.} 
    \label{fig:MSC-design}
\end{figure}

An example is given in Figure~\ref{fig:MSC-design} where a MSC arises from some underlying designer function. 
We illustrate the 1-skeleton of the MSC (equivalently, the topological skeleton of its gradient field) before and after simplification.  
Notice this approach is different from (and, in some sense, complementary to) the work on conforming MSCs~\cite{GyulassyGuntherLevine2014}, which focuses on constructing MSCs that conform with a user-specified map. 
 
Related to the design of vector fields is the notion of persistence  simplification~\cite{EdelsbrunnerLetscherZomorodian2002}. 
{\tool} allows users to view and simplify Morse vector fields with the simplification feature (highlighted in Section~\ref{sec:methods}). 
This feature connects {\tool} to persistence simplification, one of the initial motivations for studying persistence. 
 
\subsection{Inverse Problem: From Barcodes to Vector Fields}

We can also use {\tool} to study the rich area of inverse problems in topological data analysis. 
Specifically, given a persistence barcode, can we reverse engineer Morse functions or Morse vector fields that give rise to the given barcode?

\begin{figure}[!h]
    \centering
    \includegraphics[width=.98\columnwidth]{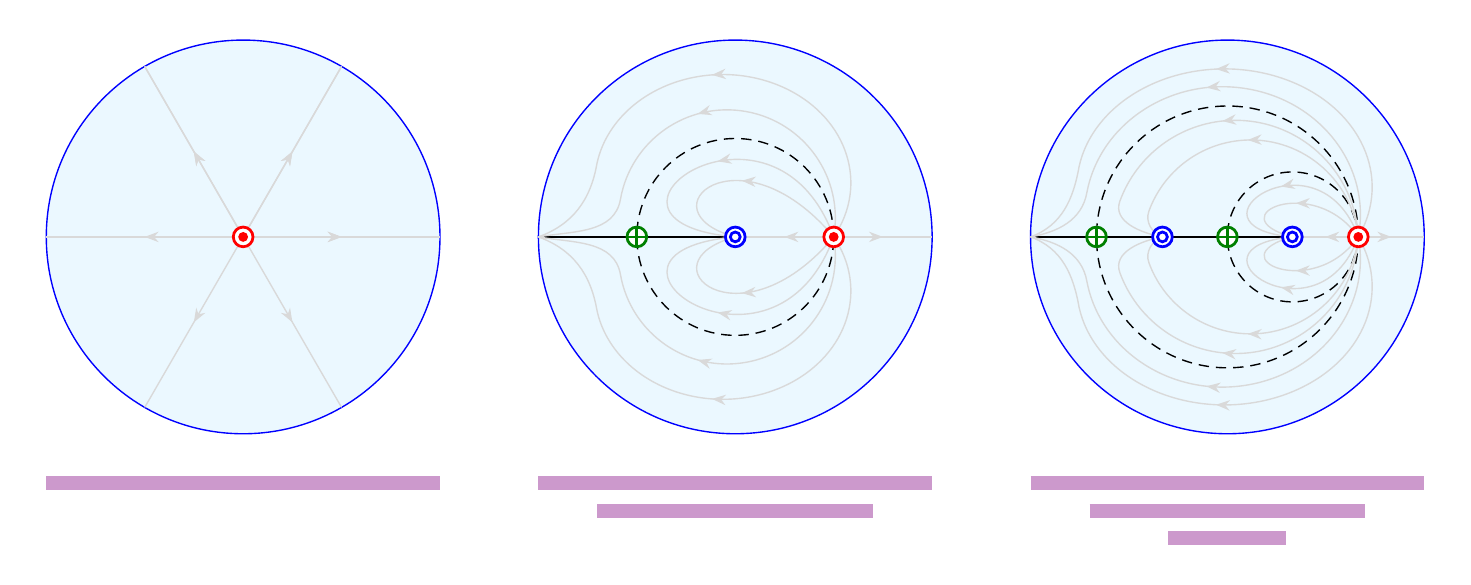}
    \vspace{-2mm}
    \caption{Morse vector field configurations, with flow visualization, that give rise to 1, 2, and 3 bars in the barcode.} 
    \label{fig:inverse-barcode}
\end{figure}

Authors in~\cite[Figure 13]{CatanzaroCurryFasy2019} studied an interesting combinatorial question: Suppose a Morse function $f$ has 6 distinct critical values and a known $0$-dimensional barcode consisting of $3$ bars nested inside each other, how many different ways can $f$ be embedded into $\Rspace^3$ while preserving the given barcode? 
Here, let us ask a simpler, but equally intriguing question: 
Given a barcode that consists of $1$, $2$, or $3$ nested bars as shown in Figure~\ref{fig:inverse-barcode}, how many Morse vector fields on the sphere can we construct that give rise to the same barcode? Assuming the largest bar is an extended persistence pair (that represents the connected component of a sphere). 

\begin{figure}[!h]
    \centering
    \includegraphics[width=.98\columnwidth]{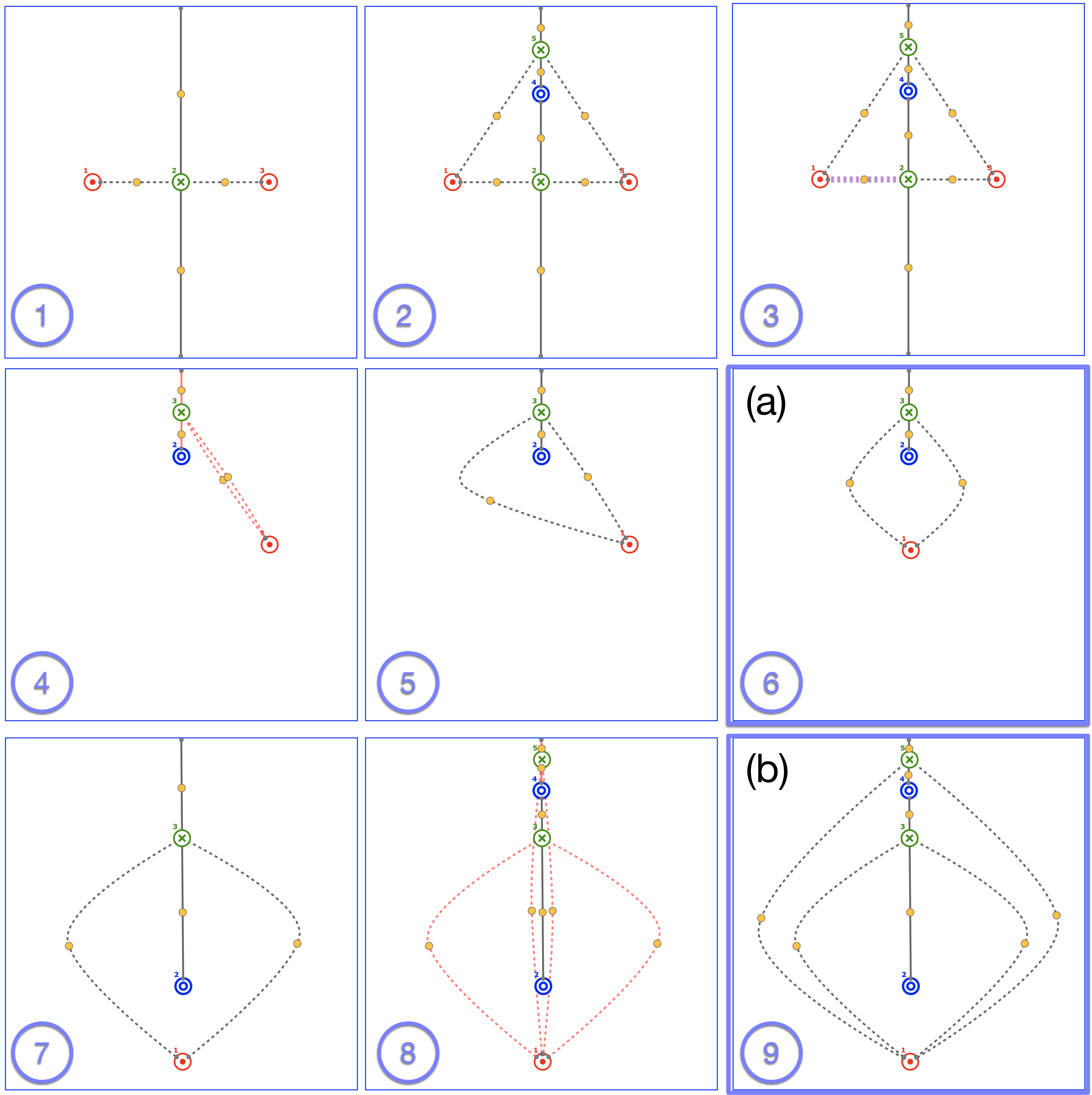}
    \vspace{-2mm}
    \caption{Two Morse vector field configurations (a, b) that give rise to two or three nested bars in a given barcode respectively.} 
    \label{fig:inverse}
\end{figure}

It turns out that there is only one valid configuration of Morse vector field that gives rise to 1, 2, or 3 bars within the above barcodes, respectively.
With {\tool},  we illustrate in Figure~\ref{fig:inverse} that we can easily construct these configurations using a few elementary moves. 
From the default configuration in Figure~\ref{fig:inverse}, we can construct configuration Figure~\ref{fig:inverse-barcode}(middle) using an edge-min move, a persistence simplification, and geometric modification. 
Moving from configuration Figure~\ref{fig:inverse-barcode}(middle) to Figure~\ref{fig:inverse-barcode}(right) involves an edge-min move and geometric modification that do not affect the barcode.
 The immediate visualization of the barcode and function adjustment options give the user direct control to construct a Morse function with the desired barcode. 

\subsection{Combinatorics of Vector Fields}
\label{sec:combinatorics}

Another inverse approach facilitated by {\tool} is the following: given a fixed number and types of critical points, how many different Morse--Smale vector fields are there with this number and type of critical points? The option to disconnect and reconnect max-saddle and min-saddle edges to and from critical points allows for an interactive search for the desired vector fields.

\begin{figure}[!h]
    \centering
    \includegraphics[width=.98\columnwidth]{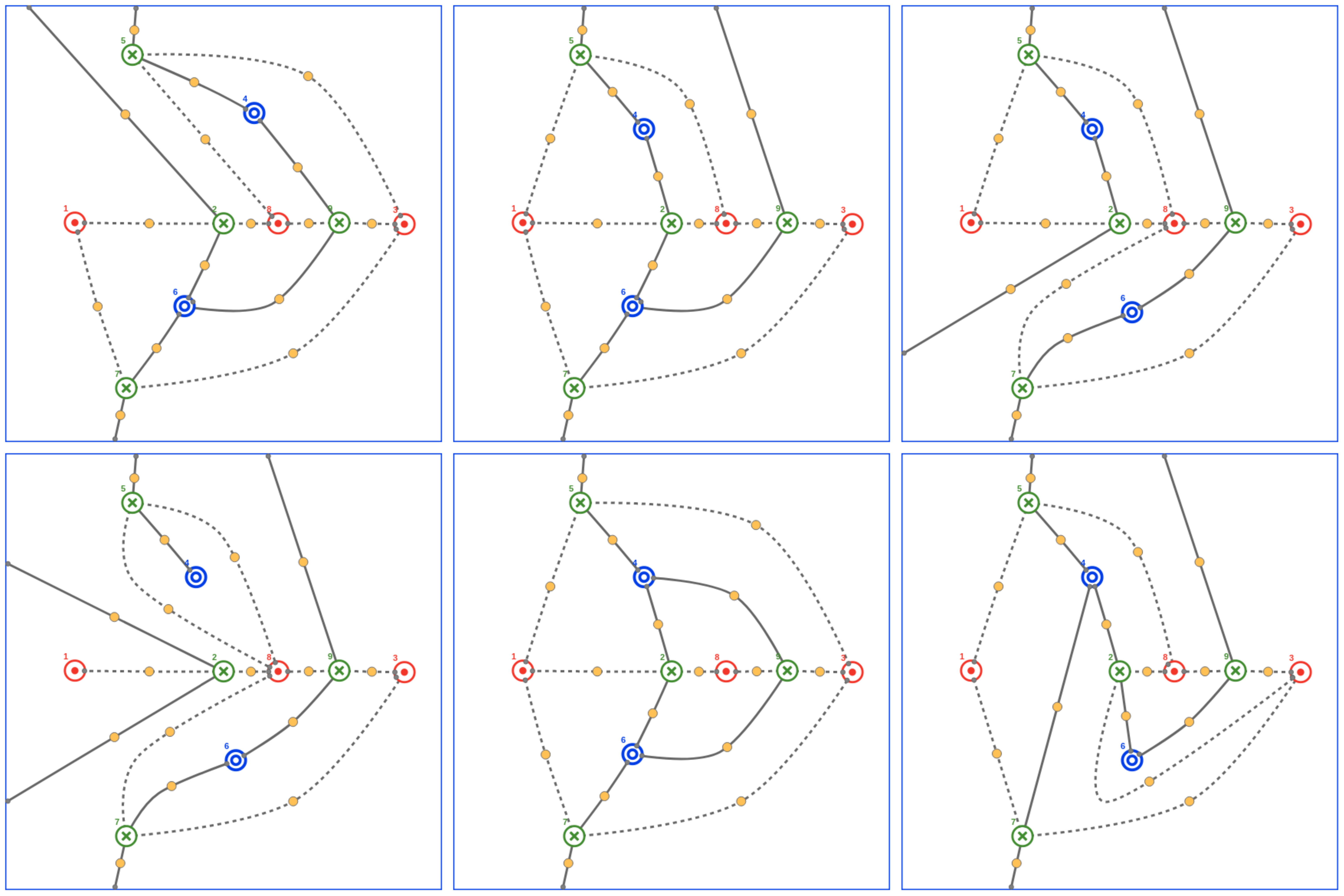}
    \vspace{-2mm}
    \caption{Vector fields as combinatorial objects.} 
    \label{fig:combinatorics}
\end{figure}

Figure \ref{fig:combinatorics} shows the use of edge connecting and disconnecting, made easier by the ``Undo" button, and ensured to be correct, rather than just abstract combinatorial objects, by the debugger. The option to save every configurations helps in more complex scenarios, as does the \texttt{json} format in which configurations are saved, which, in a future version of {\tool}, would allow for easy passing to other tools to check for graph isomorphism.

\section{Discussions}
\label{sec:discussions}
Inspired by topological data analysis, the {\tool} gives the user a tool to design and visualize Morse vector fields on the sphere. 
{\tool} also supports the design of Morse functions and Morse-Smale Complexes. 

One future direction is to support the design of general Morse--Smale vector fields by allowing periodic trajectories. 
Such an extension is nontrivial as quadrangles alone are not sufficient to describe the domain decomposition with periodic trajectories, and we can no longer use the gradient of a function to approximate the vector fields.  

Another future direction for theoretical applications is for the Wang, Fleitas, and Peixoto invariants, discussed in Section \ref{sec:topological-invariants}, to be visualized directly alongside the barcode invariant. This would allow the user to see the effect of the elementary face, edge, and vertex moves on these existing invariants, giving a unified language for working with different objects.

A possible direction for broader impact is to develop an interface with other mathematical tools, such as those that check for graph isomorphism, as mentioned in Section \ref{sec:combinatorics}, for example \texttt{Sage} or \texttt{nauty}. To help with the visualization aspect, the graph of the Morse function associated to the user-controlled Morse--Smale vector field could  also be visualized in real-time, as a 3-dimensional complement to the current 2-dimensional flow visualization. 

\newcommand{\etalchar}[1]{$^{#1}$}


\end{document}